# "Structuration" by Intellectual Organization: The Configuration of Knowledge in Relations among Structural Components in Networks of Science

Loet Leydesdorff

Amsterdam School of Communications Research, University of Amsterdam, Kloveniersburgwal 48, 1012 CX Amsterdam, The Netherlands;

loet@leydesdorff.net; http://www.leydesdorff.net.

**Abstract**

Using aggregated journal-journal citation networks, the measurement of the knowledge base in empirical systems is factor-analyzed in two cases of interdisciplinary developments during the period 1995-2005: (i) the development of nanotechnology in the natural sciences and (ii) the development of communication studies as an interdiscipline between social psychology and political science. The results are compared with a case of stable development: the citation networks of core journals in chemistry. These citation networks are intellectually organized by networks of expectations in the knowledge base at the specialty (that is, above-journal) level. The "structuration" of structural components (over time) can be measured as configurational information. The latter is compared with the Shannon-type information generated in the interactions among structural components: the difference between these two measures provides us with a measure for the redundancy generated by the specification of a model in the knowledge base of the system. This knowledge base incurs (against the entropy law) to variable extents on the knowledge infrastructures provided by the observable networks of relations.

**Introduction**

Knowledge can be considered as a meaning that makes a difference in terms of a code of communication developed within a system of relations. The code can be embodied, as in the case of an individual, or it can be reproduced—subsymbolically—in a network of distributed relations. In the latter case, *discursive* knowledge can be developed at the network level. While it is common to consider agents as knowledgeable, the concept of knowledge stored in or processed by networks requires explanation.

The knowledge carried by a network is more than and different from the sum of the knowledge carried by individual agents. For example, codified knowledge has been considered as a *common* good in evolutionary economics (Dasgupta & David, 1994). Networks can develop as structures in different dimensions that recursively condition and enable further developments. Thus, differentiation (at each moment of time) and path-dependencies potentially involving restructuration (over time) can be expected. From an evolutionary perspective, networks of relations can be considered as the historical retention mechanisms of flows of communication through the networks. These flows of communication are structured by codes of communication (Leydesdorff, 2007).

Functional differentiation among the codes of communication enables a networked system to process more complexity (Luhmann, 1986; 1995; Simon, 1972). The functions can be expected to develop evolutionarily in terms of the structural dimensions of the networks (eigenvectors), while the networks of relations develop historically in terms of (aggregates of) relations. The knowledge-based system is constructed bottom-up, but the codes of communication feed back as a top-down control mechanism. Note that different topologies are involved: relations are discrete events in design space, but the eigenvectors span a function space with continuous dimensions (Bradshaw and Lienert, 1991; Simon, 1973). The eigenvectors can be expected to change with a dynamics different from those of the networks of observable relations. A "duality of structure" is generated because the events take place in two concurrent spaces (Giddens, 1979).

From a systems perspective, structural components in the networks can be considered as condensations of the different functions carried by a networked system. One can expect that these densities are reproduced because and insofar as they are functional. However, a knowledge-based system can be expected to entertain an overlay on top of the differentiation. The different perspectives are partially integrated at the level of the overlay by using a reflexive model. This model "structurates" the configuration of eigenvectors—with reference to other possible configurations and from the perspective of hindsight—whereas the eigenvectors provide structure to the reproduction of observable variation.

In other words, a model gives meaning to the modeled. In a networked system different models can be exchanged and discursive knowledge generated as a recursive mechanism in addition to and on top of the sum total of reflexive models at the level of each individual agent or in historical components of structure (such as organizations). Figure 1 summarizes this theoretical argument in terms of an empirical research design.

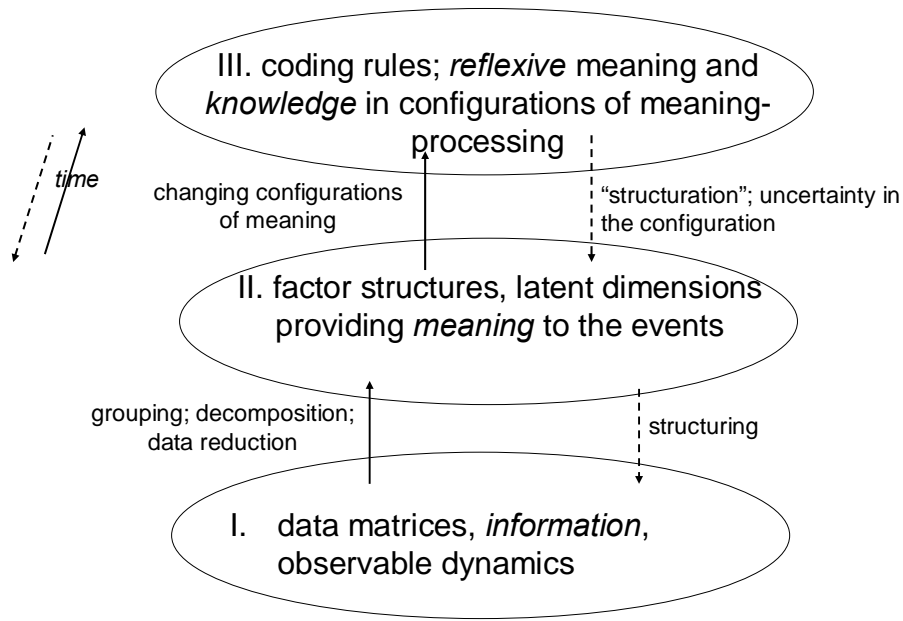

**Figure 1**: A layered process of codification of information by the processing of meaning, and the codification of meaning in terms of discursive knowledge. (Adapted from Leydesdorff (2010a), at p. 405.)

First, observable data matrices can be factor analyzed. The factor model provides structure by reducing the data. As structures develop over time, trajectories can be shaped which stabilize a system. Three selections are involved: (*i*) the momentary positioning of the data in a multidimensional space of eigenvectors, (*ii*) the positioning over time in series of events, and (*iii*) reconstruction in the present on the basis of a reflexive model (Lucio-Arias & Leydesdorff, 2009a). Whereas trajectories can develop in terms of two selections as in a process of "mutual shaping" (McLuhan, 1964), a third selection mechanism can be expected to meta-stabilize, hyper-stabilize or globalize trajectories at the regime level (Dolfsma & Leydesdorff, 2009; Dosi, 1982).

In other words, I follow Giddens's (1979, at pp. 66 ff.) distinction between *structure* and *structuration.* While structure can be operationalized in terms of latent dimensions, "structuration" governs the transformation of structures, and therefore the reproduction of a system. Giddens, however, defined a system in terms of reproduced relations, that is, as a network of observable relations (in the design space). However, the network provides only the instantiations of the system, while communication systems develop operationally in terms of different functions (Luhmann, 1995). The operations at the systems level should not be reified as network relations: the reflexive overlay does not exist as *res extensa*, but can be considered as an order of expectations in the model which potentially feeds back on the observable relations by reducing uncertainty (Husserl, 1929; Luhmann, 2002a). This additional degree of freedom enables the system to self-organize knowledge by selecting from different meanings provided to the information.[1] The model remains theoretical and therefore has the epistemological status of a set of hypotheses (Leydesdorff, 2010b).

In this study, I develop this three-layered model in empirical terms using aggregated citation relations among scientific journals as networks. Scientific journals are organized in functionally different groups. For example, articles in analytical chemistry rarely cite articles in the social sciences, or *vice versa*. Thus, one obtains densities in these networks which are reproduced from year to year for functional reasons. The densities can be considered as representations of the functions (of puzzle-solving and truth-finding) carried by the networks.

---

[1] Luhmann (1995, at p. 67) used Bateson's (1972, at p. 453) definition of information as "a difference which makes a difference." Shannon-type information is provided by a series of differences contained in a distribution, and remains meaningless before the specification of a system of reference for the measurement.

For example, some journals function to reproduce the specialty of analytical chemistry, while others reproduce sociology. Note that specialized knowledge is produced and retained at the above-journal level of journal sets in specific and knowledge-based configurations with exchange relations among them (Lucio-Arias & Leydesdorff, 2009b). The observable exchange relations provide the variation; the above-journal relations in a configuration of eigenvectors can be considered as a network of expectations.

Evolving systems develop in terms of structures and not in terms of observable (and potentially stochastic) variations. In other words, the structural components can also be considered as competing selection mechanisms on the variation. The selections provide meaning to the observable events; the orthogonal dimensions of the factor model can be used to map the different meanings in a static design. In a next step, I use the configurations among these eigenvectors as an operationalization of "structuration" and measure configurations among structural components using information theory.

Whether stabilization occurs remains an empirical question even if relations among structural components are indicated at specific moments of time. Animations enable us to visualize the resulting dynamics of the network relations. A next-order dynamics is invoked in the case of structural changes over time; this "structuration" among the components which can develop along trajectories is then changed at the regime level of the system. The reflexive model entertained in the knowledge-based system rests as a regime of pending selections on the variations, momentary selections, and historical trajectories on top of which it emerges and can be reproduced reflexively.

In other words, the events (that is, relations at the network level) are provided with different meanings by each selection mechanism. Each variable—in this study, citation pattern of a journal—is first positioned by the factor model in a multidimensional space. The factor model provides a set of momentary meanings to the variation. Additionally, the variables and eigenvectors develop over time and can be provided with historical meaning along an orthogonal (time) axis. Combinations of positional and historical meanings can be evaluated at the systems level in terms of configurations. A meaning which makes a difference at this level of a system's model can be specified as knowledge entertained by the system. The observable uncertainty in the modeled system remains the external referent of this system of expectations. If the structures in the events change over time, the system's knowledge base may be in need of an update.

**Test cases**

I focus on two instances of structural changes in network dynamics that were previously studied in detail: (1) the generation of a network of nanotechnology journals on the basis of a merger of the networks in applied physics and specific chemistry journals around 2000 (Leydesdorff & Schank, 2008), and (2) the emergence of communication studies as a network of aggregated journal-journal citation relations during the last 15 years (Leydesdorff & Probst, 2009). In these two previous studies, animations were generated for the respective fields based on trading off the stress in the representation based on multidimensional scaling at each moment of time against the stress values over time using the dynamic version of *Visone* (Baur & Schank, 2008; Leydesdorff *et al*., 2008).[2]

The animation for the nanotechnology journals (available at http://www.leydesdorff.net/journals/nanotech) first shows the embeddedness of the journal *Nanotechnology* in its relevant citation environment of journals in applied physics during the second half of the 1990s. Increasingly, chemistry journals in the environment were attracted to this focus in terms of citation relations. However, the journal *Science* played a catalyzing role in merging the two disciplinary frameworks around 2000. Thereafter, a new cluster of nano-journals emerged in which *Science* again played a role, but at this time as one of the specialist journals of the emerging field of nanoscience and nanotechnology. For example, the Institute of Scientific Information (ISI) of Thomson Reuters added the new subject category *Nanoscience & Nanotechnology* to their database in 2005. At this time, 27 journals could already be subsumed under the new category.

Communication studies—the second case—can be considered as an emerging inter-discipline between mass-communication with roots in political science and interpersonal communication rooted predominantly in social psychology. Rogers (1999, at p. 618) described this division in communication studies as "a canyon" which would be dysfunctional to the further development of the discipline. Leydesdorff & Probst (2009) focused on the delineation of a journal set that would be representative of the emerging inter-discipline.

---

[2] The dynamic version of *Visone* is freeware available at http://www.leydesdorff.net/visone.

Using the same techniques as in the study about nanotechnology, we could show that in the citation impact environment (available at http://www.leydesdorff.net/commstudies/cited) more than in the citing patterns of these journals, a third density evolved which can be identified as communication studies. Our explanation was, that despite different intellectual origins which lead to different citation patterns from other disciplinary perspectives, this third group of journals is perceived (that is, cited) increasingly as a structural component of the network. The eigenvector in the being-cited patterns of the subset of communication journals became gradually more pronounced.

In this study, the animation technique is taken one step further, first, by including the three main eigenvectors into the animations. The data is reduced to three factors because three is the lowest (and therefore most parsimonious) number of variables with interaction effects. In general, the mutual information between two variables is always positive (or zero in the case of independence), but the mutual information or, equivalently, the interaction term in a three-dimensional variance can be negative (Garner & McGill, 1956). This measure is also known as interaction information or configurational information (McGill, 1954; Yeung, 2008), and is used pervasively in many empirical sciences as a measure of interactions among three or more dimensions (Jakulin, 2005). In this study, I use it as a measure of potential synergy (in a common knowledge base) among the main components of the citation networks.

Configurational information has the seemingly attractive property of indicating synergy in the information transfer in terms of negative and positive values. However, this information is not a Shannon-measure and therefore has remained difficult to interpret (Watanabe, 1960; Yeung, 2008, at p. 59). Garner & McGill (1956, at p. 225) noted that a negative interaction term in the variance can only be the result of non-orthogonality in the design. Recently, Krippendorff (2009a, at p. 200; cf. Krippendorff, 1980) argued that circular relationships among the components are then deemed possible, which contradicts Shannon's assumptions of linear relationships. In Shannon's (1948) theory, the reception of a message cannot feed back on the message sent.

In a further elaboration (Krippendorff, 2009b), configurational information ($Q$) was identified as the net result of the Shannon-type information flow in the interactions ($I$) diminished with redundancy ($R$) in the model specification of these interactions at a next-order systems level. Krippendorff (2009a and b) considered this next-order level as an "observer," but one should keep in mind that this "observer" is only able to specify a model in terms of expectations. This "observer" thus can also be considered as a discourse. Note that the redundancy ($R$) and, therefore, the configurational information ($Q$) are *not* a property of the multivariate probability distributions in the modeled system, but their values are contained in them and can be derived from them algorithmically as (potentially negative) expected information.

In other words, because of the contextualization of the relation by a third variable, the uncertainty in the relation between two variables can be changed (as in the case of partial correlation coefficients). Krippendorff (2009b) distinguished the additional three-dimensional term using the Shannon-type decomposition ($I_{ABC \rightarrow AB:AC:BC}$) from the configurational information ($Q$) and from the redundancy ($R$) originating from the specification, and derived: $R = I - Q$. One can measure both $I$ and $Q$ in three or more dimensions of the data.

While equally uneasy about the interpretation of configurational information (as not a Shannon measure), Sun & Negishi (2008) compared this indicator with partial correlation coefficients in an empirical study of Japanese trans-sectoral (university, industry, government) and international coauthorship relations (Leydesdorff & Sun, 2009; Sun *et al.*, 2008). I shall explore this alternative measure as another indicator of configurational effects in addition to mutual information in three dimensions and Krippendorff's ternary information term. In summary, this study tests the model of knowledge generation depicted in Figure 1 against the background of two previous studies about the observable behavior of the journal systems under study.

In a third part of the empirical study, I compare the results for the two case studies with a case of relatively stable development using the ego-network of citations to the *Journal of the American Chemical Society* (*JACS*) above a certain (1%) threshold level. This data was studied in previous research projects (Leydesdorff, 1991; Leydesdorff & Bensman, 2006). In this relatively stable case, the relation between the development of structure *versus* system—that is, Giddens's (1979) "duality of structure"—can be shown to operate differently from the two cases of interdisciplinary reorganization.

**Methods and data**

Data was harvested from the CD-Rom versions of the *Journal Citation Reports* of the *Science Citation Index* and the *Social Science Citation Index* combined. In the case study about nanotechnology, all journals contributing to the citation impact environment of the journal *Nanotechnology* to the extent of 0.1% or more were included in the analysis in each year. In the case of communication studies, journal selection was based on the three ISI Subject Categories of "Communication," "Political Science," and "Social Psychology" combined with a Boolean OR-statement.[3] As noted, social psychology and political science can be considered as the two parent disciplines for the emerging inter-discipline of communication studies. Thirdly, in the case of using *JACS* as a seed journal for a relevant citation impact environment, one percent of this journal's total citations are used as a threshold for generating a citation network among approximately 20 (citing) chemistry journals in each consecutive year (1994-2007).

The citation matrices are factor-analyzed in SPSS (v. 15) using a three-factor model. The resulting factor matrices—that is, asymmetrical two-mode matrices—are used as input to

---

[3] Journals can be multiply assigned by ISI Subject Categories: on average 1.56 (± 0.76) categories/journal in 2007 (Rafols & Leydesdorff, forthcoming).

*Pajek*[4] for the visualization and to *Visone* for the animation.[2] The visualizations position the eigenvectors in the same space as the vectors using the factor loadings (that is, Pearson correlation coefficients) as (normalized) relational indicators. As a threshold, only positive correlations were included in these visualizations.[5]

The factor loadings on the three main factors can be considered as measures of association to the first three hypothesized dimensions of the multidimensional space.[6] Correlations and partial correlations between the three lists of factor loadings can be obtained directly within SPSS. In order to compute configurational information ($Q$) and Krippendorff's information measure ($I_{ABC \to AB:AC:BC}$) among the three lists of factor loadings, the (positive and negative) values are counted in bins ranging from $-1$ to $+1$ in ten steps of 0.2. This generates a three-dimensional probability distribution with $10^3$ (= 1000) cells. Dedicated software was written for the computation of $Q$ and $I_{ABC \to AB:AC:BC}$.

The mutual information in three dimensions $\mu^*$ (Yeung, 2008, pp. 51 ff.) can be calculated using Abramson's (1963, at p. 129) extension of mutual information in two to three dimensions:

$$\mu^*_{xyz} = H_x + H_y + H_z - H_{xy} - H_{xz} - H_{yz} + H_{xyz} \tag{1}$$

---

[4] Pajek is a network visualization program available at http://vlado.fmf.uni-lj.si/pub/networks/pajek/ .
[5] Because the dynamic algorithm in *Visone* uses non-metric multidimensional scaling, negative values cannot be distinguished from positive ones. The use of the value $r = 0$, however, is also convenient as a threshold (Egghe & Leydesdorff, 2009).
[6] Factor scores are by definition independent since they represent the projection of the vector on the orthogonal eigenvectors.

Each of the terms in this formula represents a (Shannon) entropy: $H_x = -\sum_x p_x \log_2 p_x$, $H_{xy} = -\sum_x \sum_y p_{xy} \log_2 p_{xy}$, etc., where $\sum_x p_x$ represents the probability distribution of attribute $x$ and $\sum_x \sum_y p_{xy}$ the probability distribution of attributes $x$ and $y$ combined. The two-dimensional transmission or mutual information ($T_{xy} = H_x + H_y - H_{xy}$) is zero in the case of two independent distributions, but otherwise necessarily positive. The resulting value of the information measure $\mu^*$ (Eq. 1) can be positive or negative depending on the relative weights of the uncertainties involved.

McGill & Quastler (1955, at p. 89) proposed calling this measure with the opposite sign a function of partial relatedness $A$ ($= -\mu^*$) because "negative interaction information is produced when the information transmitted between a pair of variables is due to a regression on a third" (McGill, 1954, at p. 108). The measure is used throughout the literature with both signs: Yeung's (2008, at pp. 51 ff.), aware that this is not a Shannon measure, proposed formalizing the mutual information in three (or more) dimensions as the information measure $\mu^*$. Krippendorff (2009a and b) followed McGill's (1954) notation, but used $Q$ instead of $A$. I follow Yeung's (2008) and Krippendorff's (2009a) notations, and hence $Q = -\mu^*$.

Figure 2 provides a metaphorical representation of this information measure based on set theory, which may nevertheless be helpful (Abramson, 1963, at pp. 130f.). If the configurational information $\mu^*$ is positive (left-hand picture), the third system $z$ receives the same information in the overlap from both $x$ and $y$. Jakulin (2005) proposed considering this as a redundancy as opposed to a synergy in the right-hand figure.

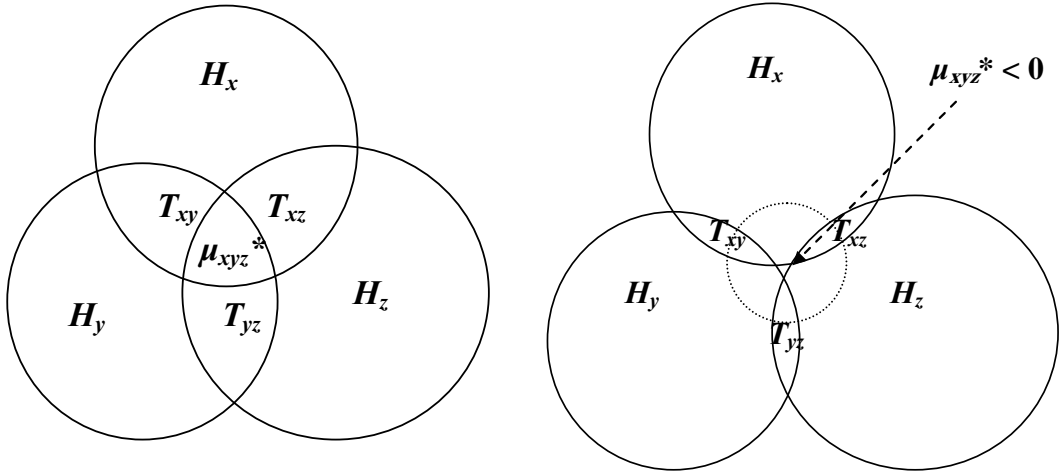

**Figure 2:** Relations between probabilistic entropies (*H*), transmissions (*T*), and configurational information ($\mu^*$) for three interacting variables.

In the right-hand case, the contextualization of the relation between *x* and *y* by *z* allows for the transmission of information via the third system in addition to the direct transmission ($T_{xy}$) between *x* and *y*. Thus, the capacity of the channel is changed because of the specification of the model. Krippendorff (2009b) proposed considering this additional capacity as a redundancy *R*: uncertainty in the system is reduced by the model specification (by an "observer"—represented here as a dotted circle), but as a feedback term.

From this perspective, the overlap in the left-hand picture adds ternary Shannon-type information ($I_{ABC \rightarrow AB:AC:BC}$) which cannot be reduced to its three binary information contents. $Q\ (=-\mu^*)$ measures the difference between the redundancy specified by the model at the systems level and the Shannon-type information generated by the interaction. The redundancy (*R*) is generated by loops in the next-order systems layer. Krippendorff (2009b, at p. 676) noted that "*interactions with loops entail positive or negative* *redundancies, those without loops do not*. Loops can be complex, especially in systems with many variables."

A technical complication is the sign of $Q$ or $\mu^*$. Yeung (2009, at p. 59) noted that one has to be cautious in referring to this information measure as a *signed* measure instead of a measure (because the latter can assume only nonnegative values). In my opinion, a negative value of $\mu^*$ in bits already indicates a redundancy; a positive value of $\mu^*$ adds to the uncertainty. The inversion of the sign between $\mu^*$ and $Q$ may easily lead to confusion about what can be considered as reduction or increase in uncertainty. For example, Krippendorff (2009b, at p. 676) formulated: "With Q(ABC) = -1, redundancy measures R(AB:AC:BC) = 1 bit, which accounts for the redundant binary interaction in AB." A redundancy of 1 bit, however, would be equal to *minus* one bit when measured as information because adding to the redundancy reduces uncertainty at the systems level.

In other words, if $I = 0$ then $R = Q$ because both $R$ and $Q$ are both defined as redundancies. Hence, $R = I + Q$ or, more precisely, the value of $R$ (as a redundancy) = $I - \mu^*$ when the latter two terms are both *measured* in bits of information. When $\mu^*$ is measured as negative, this can be considered as an imprint—in this case, remaining redundancy—generated by a modeling system. A modeling system generates redundancies by enlarging the number of possibilities and thus the maximum entropy.

The model can be considered as specified by an observer in first-order cybernetics or by a system observing itself in second-order cybernetics (e.g., Von Foerster, 1982). In the latter case, the next-order level can perform like a hyper-cycle, as indicated in Figure 2 with a dotted line. The hyper-cycle enables the system to observe the expected information content from all (orthogonal) perspectives, and thus to integrate a model without reducing the complexity to a single representation (as in the left-side picture). However, the resulting model operates with a potentially negative feedback on the necessarily positive generation of Shannon-type information.[7] If the negative feedback term prevails, self-organization is indicated as an endogenous reduction of uncertainty in the system.

Ulanowicz (1986, at pp. 142 ff.) first proposed using this potentially negative value of mutual information in three dimensions as an indicator of self-organization, that is, the net result of forward information processing and the modeling of this information processing at a next-order level within a system (Leydesdorff, 2009b). If a model is generated *within* a system as in an anticipatory system (Rosen, 1985; Dubois, 1998; Leydesdorff, 2009a) or autopoietically (Maturana, 1978; Maturana & Varela, 1980), this model provides meaning to the history of the system from the perspective of hindsight, that is, against the arrow of time. This potentially reduces uncertainty within the system, but as a negative component in an otherwise increasing uncertainty. The next-order level can be that of an external (super-)observer or a set of models using different perspectives entertained in and by a networked system. In my opinion, discursive knowledge—the empirical subject of science studies—can be considered as a prime example of knowledge entertained at a network level.

---

[7] The second law of thermodynamics holds equally for probabilistic entropy, since $S = k_B H$ and $k_B$ is a constant (the Boltzmann constant). The development of $S$ over time is a function of the development of $H$, and *vice versa.*

The specification of Krippendorff's (2009a) ternary information interaction term $I_{ABC \rightarrow AB:AC:BC}$ in bits of information can be achieved by comparing the system's state to the maximum entropy of the probability distribution. With his kind assistance I was able to reproduce Krippendorff's (1986, at p. 58) algorithm for the computation (cf. Krippendorff, 2009a, at p. 200). This routine is available at http://www.leydesdorff.net/software/krippendorff/index.htm.[8] The algorithm was further extended from the binary case to the decimal one. In other words, I used the algorithm on the same probability distribution of 10 x 10 x 10 (= 1000) probabilities as was used for the computation of the configurational information. Both $\mu^*$ ($= -Q$) and $I_{ABC \rightarrow AB:AC:BC}$ are expressed in bits. (When $\mu^*$ is an entropy, $Q$ ($= -\mu^*$) is a redundancy.)[9] Therefore, the $R$ of the model can also be expressed in bits of information.

In summary:
- I use time series of aggregated journal-journal citation networks in three cases: (1) nanotechnology, (2) communication studies, and (3) chemistry in order to indicate changes in the (factor) structures of the matrices as indicators of disciplinary and/or interdisciplinary developments;

---

[8] Krippendorff's original program (in Fortran) can be retrieved from http://www.pdx.edu/sysc/research-discrete-multivariate-modeling.

[9] $Q$ can be generalized for any dimensionality as:
$$T(:\Gamma) = \sum_{S \subseteq \Gamma} Q(S)$$
whereas mutual information can be expected to change signs with odd or even numbers of dimensions (Krippendorff, 2009b, at p. 670). In the case of three dimensions—on which we focus below as the simplest case—$Q$ is equal to the negative of mutual information in three dimensions, which will be denoted as $\mu^*$ following Yeung (2008).

- The three main components of the matrices are projected in the vector space of the journals using animations which enable us to visualize the development of interdisciplinarity in these networks;
- "Structuration" among the three components over time will be assessed using: (1) partial correlation coefficients (Sun & Negishi, 2008), (2) Krippendorff's (2009a) measure of ternary interaction information ($I_{ABC \rightarrow AB:AC:BC}$), and (3) the mutual information in three dimensions which can indicate a (potentially remaining) redundancy (Leydesdorff, 2010c).

The research question is whether and how a synergy at the above-journal level can be measured as structural change in the field(s) of science under study?

**Results**

While the above mentioned animations of the networks among journals allow us to visualize the emergence of new structural components, the animations with the eigenvectors embedded in these networks enable us to appreciate changing configurations among the components. The animations for the two fields under study with the eigenvectors embedded are brought online at http://www.leydesdorff.net/eigenvectors/nanotechnology and http://www.leydesdorff.net/eigenvectors/commstudies, respectively.

The evolution of structures in the bi-modal factor matrices are represented in two colors: green for the eigenvectors and red for the variables, that is, the aggregated citation patterns of the journals that form the networks. In the animation of the group of nanotechnology-relevant journals, journals with "nano" in their title are indicated in blue, while the node representing the journal *Science* is colored pink. In the animation of journal relations in the environment of communication studies, the 28 journals that were attributed to communication studies in 2007 by Leydesdorff & Probst (2009) are colored blue so that one can follow the emergence of this cluster.

*a. Nanoscience and nanotechnology*

The animation of the eigenvectors indicates a reorganization of structural components during the period under study. When the journal *Nanotechnology* entered the database in 1996, it was part of a structure of journals with a focus on *"Applied Physics"*. This first eigenvector relates to a second one which we designated as *"New Materials"* because in addition to chemistry journals, journals in the life sciences also load on this factor. The third factor is not easy to designate in this year (1996), but is also firmly embedded in the physics domain.

From 1997 onwards, the third factor can be designated unambiguously as *"Chemistry"*. The journal *Science* takes part in this citation network, but mainly in relation to the chemistry factor. The journal *Nanotechnology* relates to *"Applied Physics"* more than *"New Materials"*. In 1999, the factors *"Chemistry"* and *"New Materials"* become increasingly related. *Science* relates positively to all three factors, and *Nanotechnology* has shifted to a position more central in the map, by relating also to *"New Materials"*.

In 2000, the relations among the disciplinary fields are reorganized; both *Science* and *Nature* participate in this reorganization. This leads to a much closer connection between *"Applied Physics"* and *"New Materials"*, while the journal *Nanotechnology* relates both these fields to *"Chemistry"*. New journals with the root "nano" in their title emerge in the transition from 2001 to 2002, among them the journal *Nano Letters* published by the influential American Chemical Society. A triangle emerges among the three eigenvectors during the years thereafter with the nano-journals located centrally within it. The factor *"New Materials"* remains more closely related to *"Applied Physics"* than to *"Chemistry"*.

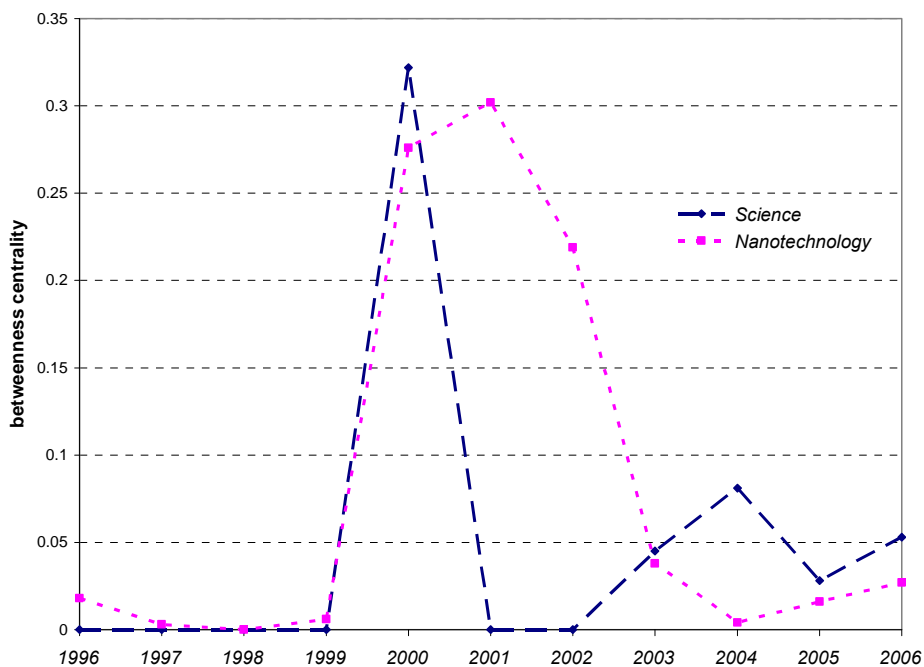

**Figure 3**: Betweenness centrality in the vector space for the journals *Science* and *Nanotechnology* during the period of the transition (cf. Leydesdorff & Schank, 2009, at p. 1816).

Leydesdorff & Schank (2008) provided a similar account of this development at the level of journals, but not of fields. The transition was indicated (*ibid.*, at p. 1816) by the increasing and decreasing betweenness centrality of the seed journal *Nanotechnology,* which peaked in 2001. In Figure 3, betweenness centrality of *Science* is added to the graph, with a peak in 2000. *Nanotechnology* took the role at the interface over from *Science* in 2001. As noted, in the years thereafter other journals were published in this same field. Would one be able to indicate the restructuration among the disciplines as taking place in 2000 using an operationalization in terms of relations among latent eigenvectors at the field level?

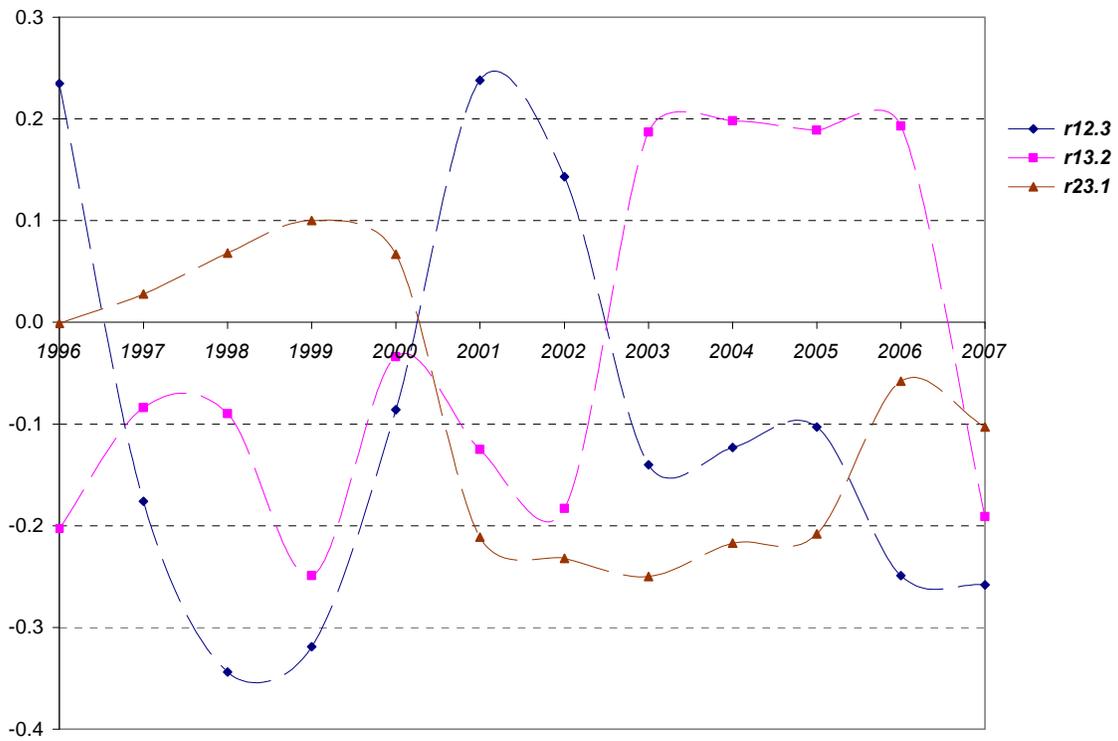

**Figure 4**: Partial correlation coefficients among the three main factors in the case of nanotechnology.

Figure 4 shows the development of the partial correlations coefficients among the three factors during the decade under study. As noted above, the factor designation is not always the same among these first three categories, but here the focus is on how the reorganization among them is represented. The reorganization is indicated as a reorganization of the three partial correlation coefficients between 2000 and 2001. The configuration remains unstable in the two years thereafter, but seems to gain more stability from 2003 onwards. The change in the position of *Science* in 2000 can be evaluated as a non-structural variation from this perspective: the development at the level of journals did not yet affect the factor structure in 2000, but did so by 2001.

The partial correlation coefficients are significantly correlated to the Pearson correlation coefficients ($r = 0.948$; $p < 0.01$). Actually, the two figures would be virtually similar, but using the Pearson correlation coefficients, the emphasis in the reorganization shifts from the first crossing of values between 2000 and 2001 towards the second one between 2002 and 2003. This result supports Sun & Negishi's (2008) argument for using the partial correlation coefficients.

Let us turn to the information measures where this difference between structure and system can be defined as $Q (= R - I)$. Figure 6 shows the development during this period of configurational information $Q$, Krippendorff's (2009a) ternary information term $I_{ABC \to AB:AC:BC}$, and the redundancy $R$ in millibits of information. Table 1 presents the data in tabular format and includes additionally the $N$ of cases.

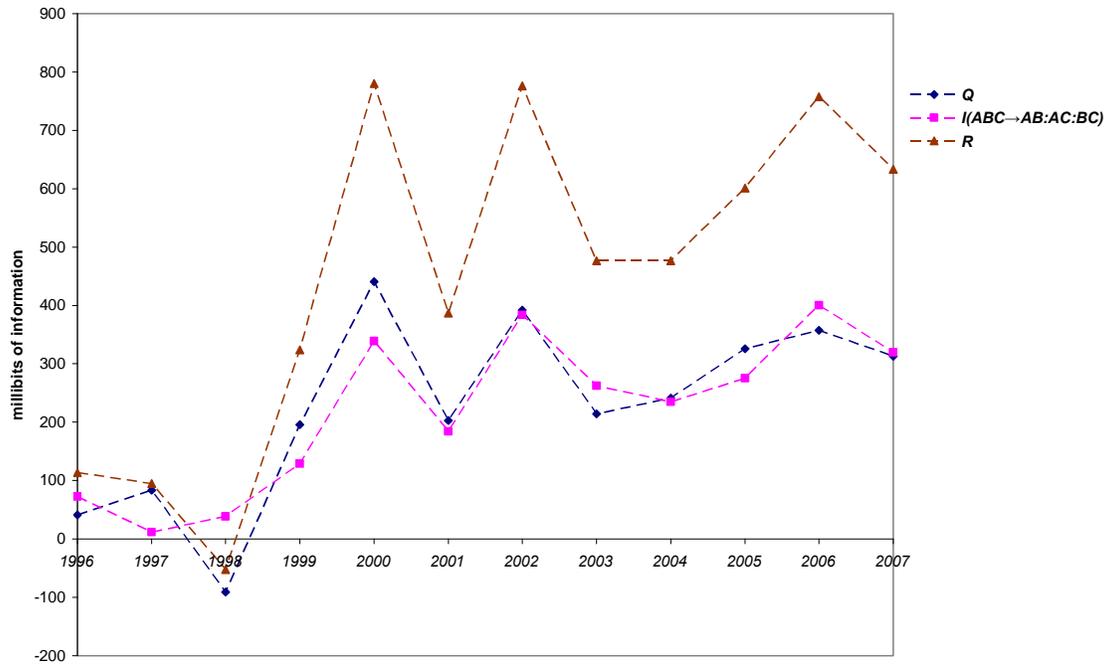

**Figure 5**: The configurational information $Q$, Krippendorff's (2009a) ternary information term $I_{ABC \rightarrow AB:AC:BC}$, and the redundancy $R$ ($= I + Q$) in millibits of information for the case of nanotechnology.

|      | $Q$ | $I_{ABC \rightarrow AB:AC:BC}$ | $R$  | $N$ |
|------|-----|-------------------------------|------|-----|
| *1996* | 41   | 73   | 114  | 41  |
| *1997* | 83   | 11   | 95   | 45  |
| *1998* | -91  | 39   | -53  | 51  |
| *1999* | 196  | 129  | 324  | 69  |
| *2000* | 441  | 339  | 780  | 72  |
| *2001* | 203  | 184  | 387  | 99  |
| *2002* | 392  | 384  | 776  | 114 |
| *2003* | 214  | 263  | 477  | 167 |
| *2004* | 241  | 235  | 476  | 172 |
| *2005* | 326  | 275  | 601  | 140 |
| *2006* | 357  | 401  | 758  | 140 |
| *2007* | 313  | 320  | 633  | 160 |

**Table 1**: The configurational information $Q$, Krippendorff's (2009a) ternary information term $I_{ABC \rightarrow AB:AC:BC}$, and the redundancy $R$ in millibits of information for the case of nanotechnology.

Figure 5 shows that both measures register the change in the configuration in 2000 with precision. The two measures are marginally different both in absolute values and in their development patterns ($r = 0.913$; $p < 0.01$), and consequently $R$ is twice as large. In other words, if $R$ is considered as the feedback term from the intellectual (self-)organization of the field surrounding the journal *Nanotechnology* as its citation impact environment, this intellectual organization is notably in disarray in 2000, but is also not stable in the years thereafter.

Perhaps, this result is a consequence of the bias introduced by focusing on a single journal and its environment. In the next study, we therefore turn to a development defined at the level of (inter-)disciplines operationalized as groups of journals in the same subject categories as defined by the Institute of Scientific Information (ISI) of Thomson Reuters.

*b. Communication Studies*

Inspection of the citation impact patterns of the individual journals (at http://www.leydesdorff.net/commstudies/cited) shows a third density increasingly emerging in addition to journals in social psychology and political science, which themselves form dense network components. A precise transition from a loose network to a structural component in the third dimension, however, is not clearly indicated. Upon visual inspection, the development seems mainly gradual. Is it possible to indicate structural change in this development using our systems measures?

In the years 1994-1996, the journal *Public Opinion Quarterly* played a key role in relating the communication studies journals first to journals in the political sciences, and then also to journals in social psychology. The years 1996-1998 witnessed notably an increase in the density of relations between communication studies and social psychology. In 1998, *Public Opinion Quarterly* and *Human Communications Research* were central to the interfaces of the emerging cluster of journals in communication studies with journals in political science and social psychology, respectively.

In terms of eigenvector development, the communication studies journals were first (1994-1995) immersed in the internal complexity of two factors (Factors Two and Three) which can both be designated as political science. One of these factors focuses on political units of analysis such as comparisons among nation states, and the other more on political processes, led by American journals (such as the *American Political Science Review* and *American Political Quarterly*). The communication studies journals load negatively on the former of these two factors, but neutrally on the latter.

In 1996, this profile is enhanced: both the *Journal of Communication* and *Communication Research*—two flagship journals of the International Communications Association (ICA)—load negatively (with –0.641 and –0.630, respectively) on a factor that is otherwise still dominated with a positive sign by journals such as the *European Journal of Political Research*, the *British Journal of Political Science*, and *Election Studies*. This third factor is a mixture of the two components in this year preceding the transition. In 1997, however, the third factor can be designated unambiguously as *"Communication Studies"* in addition to a first factor representing *"Social Psychology"* and a second *"Political Science"*. (The American journals mentioned above dominate this latter factor, but the other group is part of it given a three-factor model.)

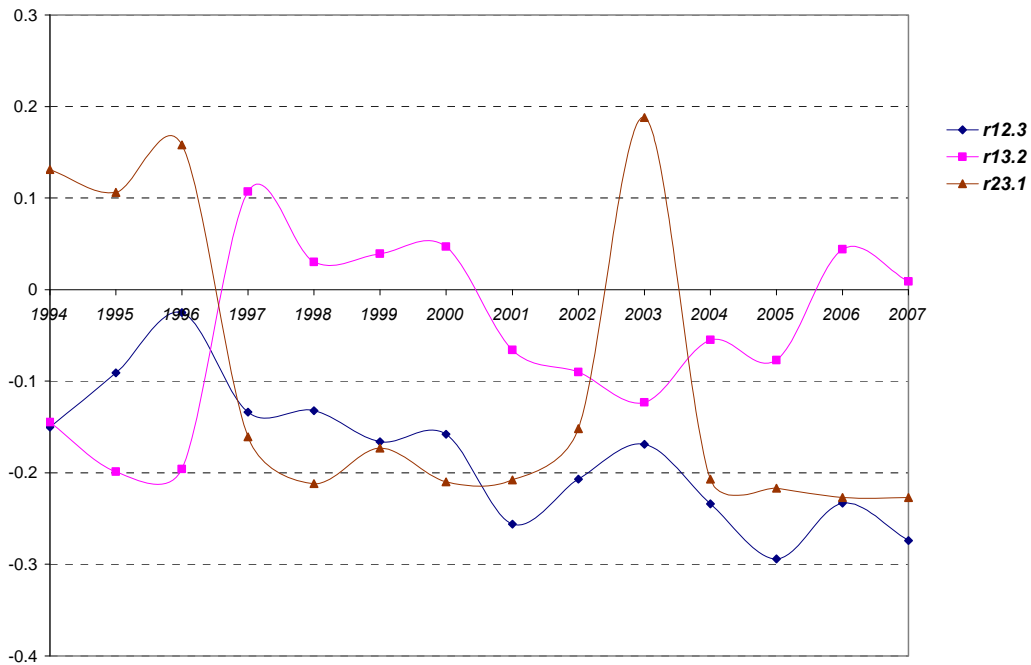

**Figure 6**: Partial correlation coefficients among the factor loadings on three main factors in the case of communication studies.

Figure 6 indicates the changes: the partial correlations of the loadings on both Factors One (social psychology) and Two (political science) with Factor Three change sign between 1996 and 1997. The third factor groups a set of journals in communication studies in the latter year for the first time. The other major event indicated, is the disappearance of the (third) communication-studies factor in 2003. In this year only, the pre-1997 configuration is restored for a single year. This effect in 2003 is also visible in the animation (at http://www.leydesdorff.net/eigenvectors/commstudies/).

The partial correlations are in this case even more strongly correlated to the Pearson correlations than in the previous one ($r = 0.981$; $p < 0.01$). The difference between the two matrices mainly exhibits the huge effect in 2003, and to a smaller extent the developments in 1997, that is, the emergence of a new cluster of communication studies journals. However, the earlier change was crucial. In other words, the partial correlation coefficients provide descriptive statistics of the events visible in the animations. However, these measures cannot provide a measure of the three-way interaction effects.

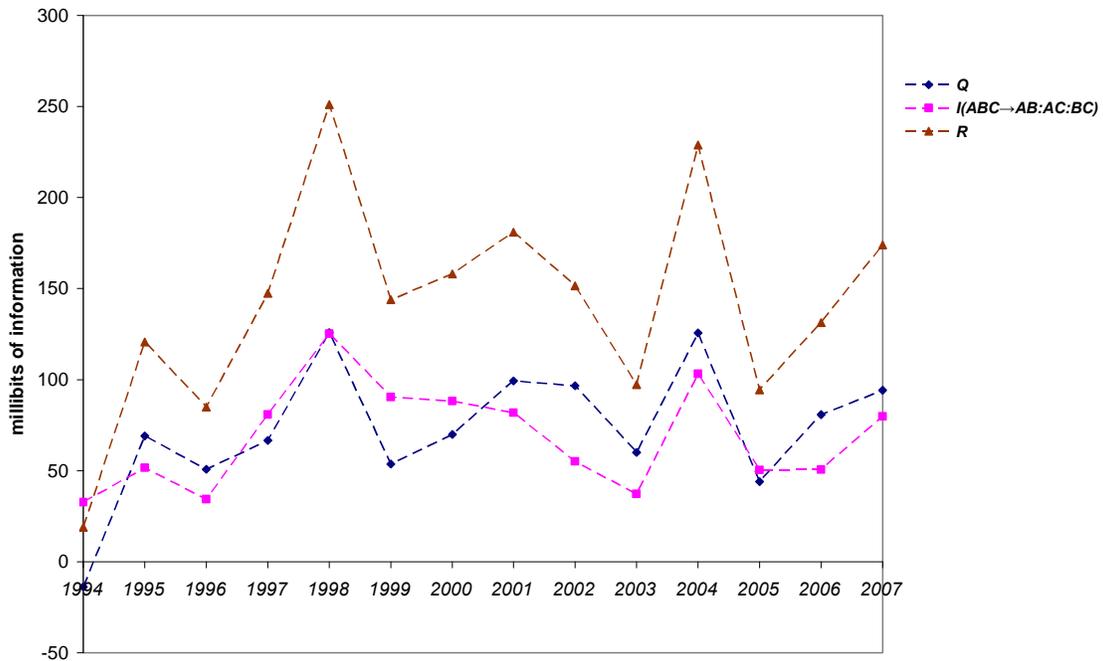

**Figure 7**: The configurational information $Q$, Krippendorff's (2009a) ternary information term $I_{ABC \to AB:AC:BC}$, and the redundancy $R$ $(= I + Q)$ in millibits of information for the case of communication studies.

|      | $Q$ | $I_{ABC \rightarrow AB:AC:BC}$ | $R$ | $N$ |
|------|-----|-------------------------------|-----|-----|
| 1994 | -14 | 33  | 19  | 122 |
| 1995 | 69  | 52  | 121 | 123 |
| 1996 | 51  | 34  | 85  | 128 |
| 1997 | 67  | 81  | 147 | 139 |
| 1998 | 126 | 125 | 251 | 144 |
| 1999 | 54  | 90  | 144 | 148 |
| 2000 | 70  | 88  | 158 | 149 |
| 2001 | 99  | 82  | 181 | 155 |
| 2002 | 97  | 55  | 152 | 158 |
| 2003 | 60  | 37  | 97  | 162 |
| 2004 | 126 | 103 | 229 | 157 |
| 2005 | 44  | 50  | 94  | 164 |
| 2006 | 81  | 50  | 131 | 168 |
| 2007 | 94  | 80  | 174 | 177 |

**Table 2**: The configurational information $Q$, Krippendorff's (2009a) ternary information term $I_{ABC \rightarrow AB:AC:BC}$, and the redundancy $R$ ($= I + Q$) in millibits of information for the case of communication studies.

Figure 7 shows the development of the configurational information $Q$, Krippendorff's (2009a) ternary information term $I_{ABC \rightarrow AB:AC:BC}$, and the redundancy $R$ in millibits of information. Table 2 provides this data in tabular format. The figure indicates the reorganization during the second half of the 1990s. Both curves peak in 1998 and 2004: the Pearson correlation coefficient between $Q$ and $I_{ABC \rightarrow AB:AC:BC}$ is 0.704 ($N = 14$; $p < 0.05$).

The latter peak represents the recovery after the disappearance of the emerging configuration in 2003, and the former the initial emergence of communication studies as a structural component in 1998. This latter year corresponds with the spanning of a triangular structure among the three factors in the animation at http://www.leydesdorff.net/eigenvectors/commstudies. Appearing as an independent (third) factor for the first time in 1997, the component representing *Communication Studies* further developed into a separate dimension of the data in 1998.

In the years after 1998, the emerging configuration remains volatile. As noted above, the factor solution for 2003 shows a pattern similar to that before 1997. Indeed, the curves for both *I* and *Q* show a low for this year, with higher values for 2004. Leydesdorff & Probst (2009) noted the further development of a group of journals about *Discourse Analysis* in 2006 and 2007 on the basis of a more detailed factor analysis in six dimensions.

Perhaps one can expect a different relation between the historical generation of Shannon-type information (*I*) and redundancy (*R*) generated by the model in more stable fields of science; this may lead to larger differences between *I* and *Q*. In these two case studies, however, the focus was on rearrangements in the structures and how these are indicated by *Q* and *I*. It seems that both *Q* and *I* can be used because the two indicators are correlated in the case of changes at the systems level. How might this be different in the case of a relatively stable configuration?

*c. The citation impact environment of the JACS*

The citation impact environment of the *Journal of the American Chemical Society* (*JACS*) can be considered as such a stable configuration (Leydesdorff, 1991). This flagship journal of the American Chemical Society was founded in 1879 and had an impact factor of 7.885 in 2007. Its mere volume of approximately 3,000 publications each year makes *JACS* the leading journal in the field of chemistry in terms of citations and references. In 2007, the citation impact environment of this journal consists of a structure of three main components, explaining 72.3% of the variance, and two smaller components which load on a fourth factor (explaining another 6.9%) with opposite signs. Table 3 provides the rotated component matrix for the four-factor solution of the journal-journal citation matrix.

|  | Component | | | |
|---|---|---|---|---|
|  | 1 | 2 | 3 | 4 |
| Tetrahedron | **.944** | | | |
| Tetrahedron Lett | **.941** | | | |
| J Org Chem | **.936** | .197 | -.106 | |
| Eur J Org Chem | **.922** | .187 | -.112 | |
| Org Lett | **.888** | .301 | -.121 | |
| J Am Chem Soc | | **.889** | | .219 |
| Chem-Eur J | .138 | **.881** | .245 | |
| Chem Rev | .295 | **.846** | | .265 |
| Angew Chem Int Edit | .123 | **.769** | | -.203 |
| Chem Commun | .212 | **.753** | .426 | -.261 |
| J Organomet Chem | -.132 | | **.845** | |
| Dalton T | -.406 | .230 | **.803** | -.103 |
| Organometallics | -.213 | .118 | **.787** | |
| Inorg Chem | -.400 | .406 | **.572** | |
| J Phys Chem A | -.190 | | -.139 | **.921** |
| J Phys Chem B | -.494 | .104 | -.579 | **.334** |
| Langmuir | -.448 | | -.576 | *-.218* |
| Macromolecules | -.354 | -.265 | -.396 | *-.335* |

Extraction Method: Principal Component Analysis.
Rotation Method: Varimax with Kaiser Normalization.
Rotation converged in 6 iterations.

**Table 3**: Four-factor solution for the citation impact environment of *JACS* in 2007.

The three major components (organic, general, and inorganic chemistry) are present in each year of *JACS*'s citation environment as the first three components, although in some years the order among them changes. In previous studies, these environments were studied both in terms of subject headings in the catalogue of the Library of Congress for validation purposes (Leydesdorff & Bensman, 2006) and in terms of their dynamic development (Leydesdorff, 1991). In sum, these three categories provide us with a relatively stable configuration of structural components.

The stability of the configuration can be illustrated with an animation using *PajekToSVGAnim*.[10] The animation is available at http://www.leydesdorff.net/eigenvectors/jacs/index.htm. This animation shows the extreme stability of the three-factor solution in terms of eigenvectors representing organic, general, and inorganic chemistry journals.

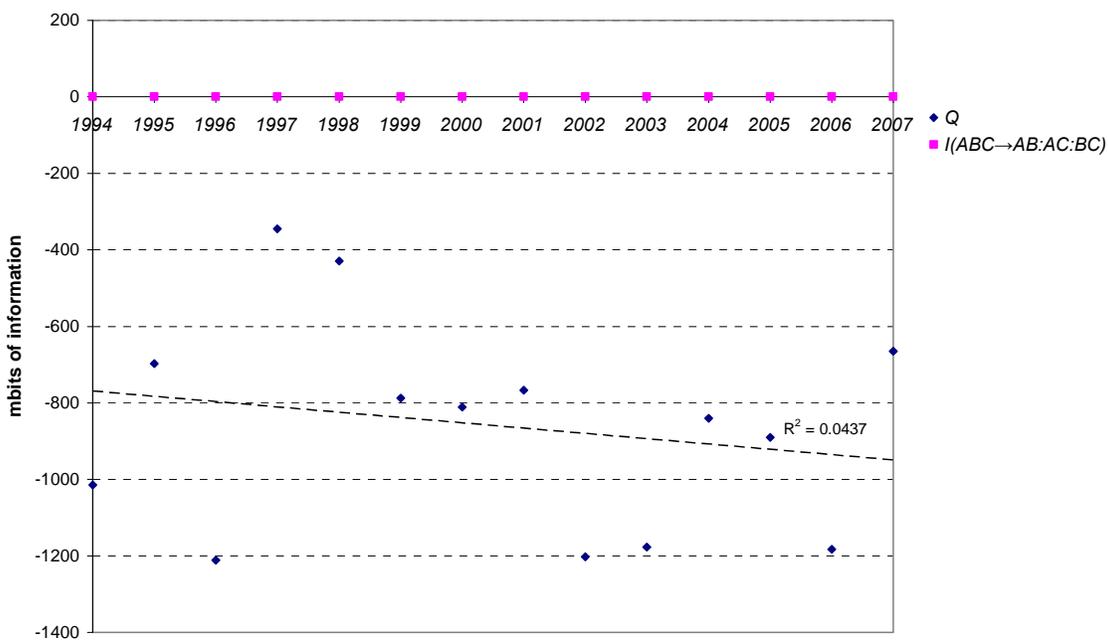

**Figure 7**: The configurational information *Q* and Krippendorff's (2009a) ternary information term $I_{ABC \rightarrow AB:AC:BC}$ in millibits of information for the citation impact environment of the *Journal of the American Chemical Society* (*JACS*); threshold = 1%.

---

[10] PajekToSVGAnim.exe is freely available for non-commercial usage at http://vlado.fmf.uni-lj.si/pub/networks/pajek/SVGanim/default.htm. Unlike *Visone* this program allows for including negative factor loadings.

I first tried to apply these same methods to the citation matrices of *JACS* using a 1% threshold. The results are shown in Figure 7. The values for $I_{ABC \rightarrow AB:AC:BC}$ are vanishingly small (less than 0.1 millibits) and the values of $Q$ are always negative. In other words, this is not a three-dimensional, but a two-dimensional structure without ternary interactions among the three main dimensions, and with variable values of the mutual information in two dimensions.[11] The general chemistry journals function in this environment as an overlapping interface between organic and inorganic chemistry journals. This interface function, however, varies from year to year.

If one extends the analysis to the 160+ journals participating in the citation impact environment of *JACS* at the 0.1% level, the journals in physical chemistry form a third group, and intellectual organization among the three dimensions of this system can now be expected. Figure 8 shows the results. On the right-hand side, I added the same analysis using the approximately 115 journals which constitute the environment not in terms of cited patterns, but citing—at the same 0.1% threshold level—because I expected intellectual organization to be more pronounced when using the citation behavior of the authors in these leading chemistry journals than in the cited direction. This is indeed the case.

---

[11] The difference in the sign is generated because $Q$ is computed assuming three dimensions.

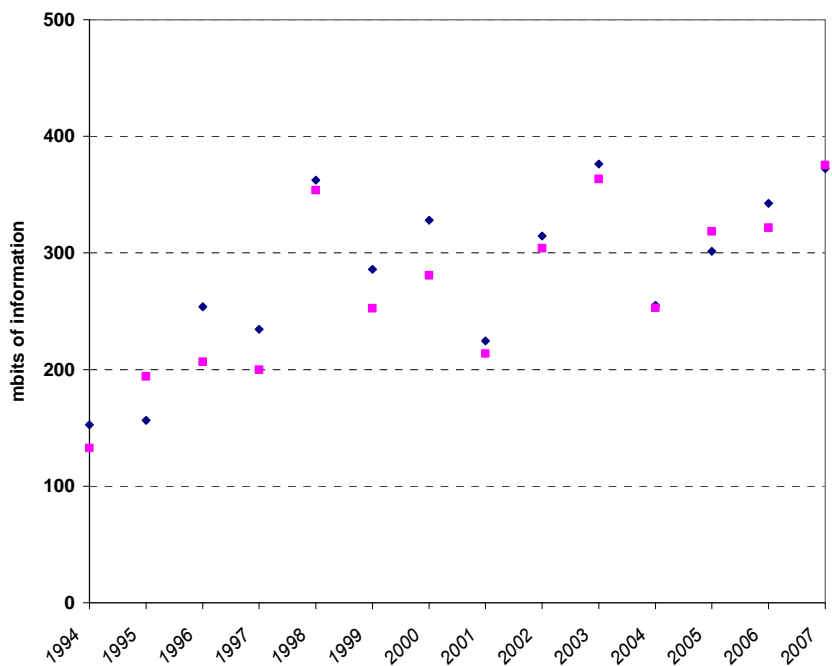 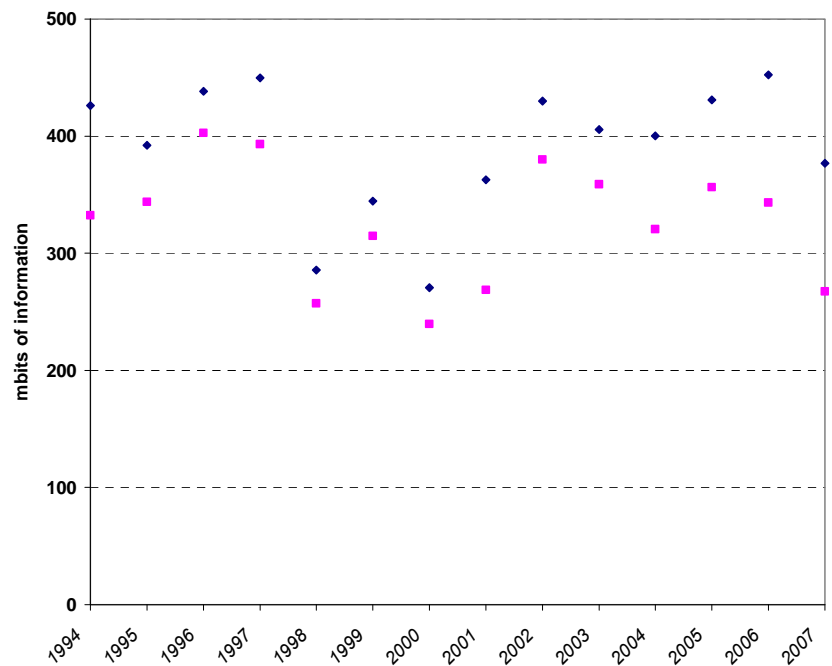

**Figure 8**: $Q$ (♦) and $I_{ABC \rightarrow AB:AC:BC}$ (■) values for the cited (left) and citing patterns (right) in the citation environment of *JACS* in terms of the three main dimensions: organic, inorganic, and physical chemistry; threshold > 0.1%.

The remarkable finding is again the high correlation between the values for $I_{ABC \to AB:AC:BC}$ and $Q$, both cited ($r = 0.95$; $p < 0.01$) and citing ($r = 0.86$; $p < 0.01$). However, the values in the two directions of cited *versus* citing are negatively correlated ($r = -0.29$ for $Q$ and $r = -0.28$ for $I_{ABC \to AB:AC:BC}$; *n.s.*). While these values increase in the cited direction, they are relatively stable in the citing direction, albeit with a low in the years 1998-2001 for both values. The relatively high values—when compared with the two previous case studies—can perhaps be explained by the specific role of general-chemistry journals (such as *JACS*) which exhibit inter-factorial complexity by loading on all three components. These journals intellectually organize the field at a level above the specialties.

In summary, this third case teaches us that organic and inorganic chemistry are strongly interwoven in terms of their intellectual organization—in which this journal (*JACS*) serves as an "observer." This co-evolution between two dimensions does not provide us with ternary interaction information, but mutual information. By extending the scope to physical chemistry, a continuous reorganization and reproduction of the relations among the three fields in terms of citation relations seems indicated. The general chemistry journals serve this mechanism of integration and accordingly reproduce the differentiation.

**Conclusions and discussion**

Before turning to the theoretical conclusions, let me first summarize the empirical findings:

1. Using the Ego-networks of the journal *Nanotechnology*—which was entered into the database in 1996—the emergence of nanotechnology as an interdisciplinary development could be indicated in terms of journal-journal citation in 2000. Note that this is before the major priority programs were put in place, given that citation is a delayed indicator. The development remained turbulent thereafter although some stabilization of the field as an interdiscipline occurred during the years 2001-2003;

2. Using field definitions of the collections of journals relevant for communication studies, social psychology, and political science, the emergence of communication studies as an interdiscipline and then increasingly an independent field of studies could be traced as emerging in 1997, and then relatively stabilized in the years thereafter. In 2003, the configuration was meta-stable during one year in terms of aggregated citations. The gain in identity of the set is more pronounced in the cited than the citing dimension because in the latter patter the orientation to the mother disciplines is still strong;

3. Using the Ego-networks of the *Journal of the American Chemistry Society* (*JACS*), the two interdisciplinary disturbances and developments could be compared to developments in a relatively stable field of studies. The two subdisciplines of organic and inorganic chemistry are interfaced by general chemistry journals such as *JACS* more than that a synergy is generated as a redundancy. If "physical chemistry"

The results in the two cases of interdisciplinary developments suggest that both $I_{ABC \rightarrow AB:AC:BC}$ and $Q$ provide us with indicators of change in configurations among structural dimensions. Conceptually, however, these two measures are very differently defined. Whereas $I_{ABC \rightarrow AB:AC:BC}$ indicates Shannon-type information caused by the three-way interaction, $Q$ is the complement between this historical uncertainty and the redundancy provided by the model. Since the model provides meaning to the historical events, one could also consider $Q$ as a measure of meaningful information, that is, the difference between (Shannon-type) information and its meaning for a receiving system (e.g., an observer). Brillouin (1962) noted that meaningful information can also be negative and proposed the terminology of "negentropy" for meaningful information (cf. information as "a difference which makes a difference" [Bateson, 1972, at p. 489]).

In the third case of stable disciplinary development, $Q$ was strongly negative and the historical interaction among the components ($I_{ABC \rightarrow AB:AC:BC}$) vanished. In this case, the observable network relations did not affect the interactions among the three components historically, but the information remained reflexively meaningful for the reproduction of the system as a knowledge-based configuration. Since $I$ and $Q$ are both high in the case of interdisciplinary developments (Figures 5 and 7), not only was uncertainty produced within the system, but this information was also meaningful at the systems level.

(Partial) correlation coefficients among the structural dimensions provided us with descriptive statistics of changes. The latter could also be visualized by positioning the eigenvectors among the variables, that is, by using the rotated factor matrices as input to the animations. Insofar as one can observe an increase (or decrease) in complexity by using these animations, this has to be considered as Shannon entropy, since $Q$ provides a difference which cannot be observed directly. The value of $Q$ is an effect of the configuration which provides us with an algorithmic access (Equation 1) to the model generating redundancy. This model can be entertained by an external observer in the case of first-order cybernetics or an observing subroutine of the system. In the latter case, the theoretical frame of reference can be provided by the theories of both anticipatory systems (Rosen, 1985) and *autopoiesis* (Maturana & Varela, 1980; cf. Leydesdorff, 2009).

In other words, a model is offered for how knowledge can be generated and self-organized in networks. Beyond being generated, discursive knowledge can again be communicated in the knowledge networks of social systems. Thus, the next-order level can be considered as an overlay which loops back into the information processing (Maturana, 2000). The order of expectations coevolves with the order of events in a knowledge-based system. In my opinion, the reflexivity of human agency drives the loop because the expectations have to be articulated into new knowledge claims. The distribution and communication of the latter provide the variation on which the different selection mechanisms can operate. Note that the development of discursive knowledge presumes the flexibilities of human language and reflexivity (Giddens, 1984; Leydesdorff, 2000; Luhmann 2002b). Both recursions (with the arrow of time) and incursions (against the arrow of time) are involved (Dubois, 1998).

This model captures Giddens's (1979) concept of "structuration" and provides it with an empirical operationalization. Furthermore, this concept could be positioned with reference to Luhmann's (1995) social systems theory and Maturana and Varela's (1980) theory of *autopoiesis*. The mechanism for reproduction of structure in networks is different from—orthogonal to—the network structure itself. Structure is static and (re)produced at each moment of time. Giddens's dictum that "structure only exists as 'structural properties'" accords with the factor-analytic model: eigenvectors can be considered as structural components of a network.

The configuration among the hypothesized dimensions can be entertained as a model of structure by a knowledge-based system. Because the model is only available reflexively (that is, in terms of expectations), structuration should not be reified: it operates as a "duality of structure" but in a virtual domain (Giddens, 1979, pp. 81 ff.). This duality was specified in terms of Shannon-type information aggregated into structure *versus* the redundancy generated by the model. $Q$ measures the difference between these counteracting dynamics, that is, the imprint of the (self-)organization at the systemic level on the historical development of structures. The structural components or eigenvectors provide the historical instantiations of structure. Systemness, however, should in this case be understood not in Giddens's (1979, at p. 66) sense as "reproduced relations," but as Luhmann's (and Husserl's) "horizons of meaning" which can be codified in the knowledge base of a system as universes of possible communications.

**Acknowledgement**

I am grateful to Klaus Krippendorff for his further specification of the algorithm for the computation of maximum entropies in the case of ternary interaction terms. Andrea Scharnhorst was a sparring partner for developing these ideas in discussions about previous drafts.

McGill, W. J. (1954). Multivariate information transmission. *Psychometrika,* 19(2), 97-116.

McGill, W. J., & Quastler, H. (1955). Standardized nomenclature: An attempt. In H. Quastler (Ed.), *Information Theory in Psychology: Problems and Methods* (pp. 83–92). Woodbury, NY: The Free Press.

McLuhan, M. (1964). *Understanding Media: the Extension of Man*. New York: McGraw-Hill.

Rogers, E. M. (1999). Anatomy of the two subdisciplines of communication study. *Human Communication Research,* 25(4), 618-631.

Rosen, R. (1985). *Anticipatory Systems: Philosophical, mathematical and methodological foundations*. Oxford, etc.: Pergamon Press.

Shannon, C. E. (1948). A Mathematical Theory of Communication. *Bell System Technical Journal,* 27, 379-423 and 623-356.

Simon, H. A. (1973). Does scientific discovery have a logic? *Philosophy of Science 40*, 471-480.

Simon, H. A. (1973). The Organization of Complex Systems. In H. H. Pattee (Ed.), *Hierarchy Theory: The Challenge of Complex Systems* (pp. 1-27). New York: George Braziller Inc.

Sun, Y., & Negishi, M. (2008). Measuring relationships among university, industry and the other sectors in Japan's national innovation system. In J. Gorraiz & E. Schiebel (Eds.), *10th International Conference on Science and Technology Indicators Vienna* (pp. 169-171). Vienna, 17-20 September 2008: Austrian Research Centers.

Sun, Y., Negishi, M., & Nisizawa, M. (2009). Coauthorship Linkages between Universities and Industry in Japan. *Research Evaluation* (in print).

Ulanowicz, R. E. (1986). *Growth and Development: Ecosystems Phenomenology*. San Jose, etc.: toExcel.

Von Foerster, H. (1982). *Observing Systems* (with an introduction of Francisco Varela ed.). Seaside, CA: Intersystems Publications.

Watanabe, S. (1960). Information theoretical analysis of multivariate correlation. *IBM Journal of research and development,* 4(1), 66-82.

Yeung, R. W. (2008). *Information Theory and Network Coding*. New York, NY: Springer; available at http://iest2.ie.cuhk.edu.hk/~whyeung/post/main2.pdf .43